\documentclass[paper,a4paper,11pt,twoside]{JHEP3}
\usepackage{graphicx}
\usepackage{amsmath}
\usepackage{axodraw}
\usepackage{cite}

\def\gev{\,\mathrm{GeV}}
\def\mev{\,\mathrm{MeV}}

\def\SU{\mathrm{SU}}

\def\kl3{K_{\ell 3}}
\def\lqcd{\Lambda_{\textrm{QCD}}}
\def\lchi{\Lambda_\chi}
\def\mkbar{\bar{m}_K}
\def\qsqmax{q^2_{\textrm{max}}}

\title{%
$\SU(2)$ chiral perturbation theory for $K_{\ell 3}$ decay amplitudes}
\author{J.M.~Flynn and C.T.~Sachrajda,\\
School of Physics and Astronomy, University of Southampton,\\
Southampton, SO17 1BJ, UK\\
E-mail: {\tt jflynn@phys.soton.ac.uk, cts@phys.soton.ac.uk}}

\author{RBC and UKQCD Collaborations}

\preprint{%
SHEP--08--26}

\abstract{%
We use one-loop $\SU(2)_L\times \SU(2)_R$ chiral perturbation theory ($\SU(2)$ ChPT) to study the behaviour of the form-factors for semileptonic $K\to\pi$ decays with the pion mass at $q^2=0$ and at $q^2_{\textrm{max}}=(m_K-m_\pi)^2$, where $q$ is the momentum transfer. At $q^2=0$, the final-state pion has an energy of approximately $m_K/2$ (for $m_K\gg m_\pi$) and so is not \textrm{soft}, nevertheless it is possible to compute the chiral logarithms, i.e. the corrections of $O(m_\pi^2\log(m_\pi^2))$. We envisage that our results at $q^2=0$ will be useful in extrapolating lattice QCD results to physical masses. A consequence of the Callan-Treiman relation is that in the $\SU(2)$ chiral limit ($m_u=m_d=0$),
the scalar form factor $f^0$ at $\qsqmax$ is equal to $f^{(K)}/f$, the ratio of the kaon and pion leptonic decay constants in the chiral limit. Lattice results for the scalar form factor at $\qsqmax$ are obtained with excellent precision, but at the masses at which the simulations are performed the results are about 25\% below $f^{(K)}/f$ and are increasing only very slowly. We investigate the chiral behaviour of $f^0(\qsqmax)$ and find large corrections which provide a semi-quantitative explanation of the difference between the lattice results and $f^{(K)}/f$. We stress the generality of the relation $f^0_{P\to\pi}(\qsqmax)=f^{(P)}/f$ in the $\SU(2)$ chiral limit, where $P=K,D$ or $B$ and briefly comment on the potential value of using this theorem in obtaining physical results from lattice simulations.}


\begin{document}

\section{Introduction}\label{sec:intro}

One of the most precise methods to extract the $V_{us}$ element of the CKM-matrix is to use $K\to\pi\ell\nu_\ell$ semileptonic decays ($K_{\ell 3}$ decays), where $\ell$ is an electron or a muon. The combination $|V_{us}f^+(0)|$ can be determined from the experimental rate
\begin{equation}\Gamma_{K\to\pi\ell\nu_\ell}=C_K^2\frac{G_F^2m_K^5}{192\pi^3}\,I\,S_\mathrm{EW}\,\left(1+2\Delta_{\SU(2)}+2\Delta_\mathrm{EM}
\right)\,|V_{us}|^2\,|f^+(0)|^2\,,\end{equation}
where $I$ is a phase space integral which can be evaluated from the experimentally determined shape of the form-factors and $\Delta_{\SU(2)}$, $\Delta_\mathrm{EM}$ and $S_\mathrm{EW}$ contain the calculable corrections due to isospin breaking, electromagnetic and short-distance electroweak effects respectively. $C_K^2=1/2\,(1)$ is the Clebsch-Gordan coefficient for the neutral (charged) kaon decay and
$f^+(0)$ is the form factor defined from
\begin{eqnarray}\langle\,\pi(p_\pi)\,|\,\bar{u}\gamma_\mu s\,|\,\bar{K}(p_K)\,\rangle &=& (p_K+p_\pi)_\mu\,f^+(q^2)+(p_K-p_\pi)_\mu\,f^-(q^2)\\
&=& \left[(p_K+p_\pi)^\mu-q^\mu\frac{m_K^2-m_\pi^2}{q^2}\right]f^+(q^2)+q^\mu \frac{m_K^2-m_\pi^2}{q^2}f^0(q^2)\,,\nonumber\end{eqnarray}
where $q$ is the momentum transfer $q=p_K-p_\pi$. At $q^2=0$ we have $f^+(0)=f^0(0)$\,. The Particle Data Group(2008)~\cite{Amsler:2008zz} quotes
\begin{equation}|V_{us}f^+(0)|=0.21668(45)\,,\label{eq:vusfplus}\end{equation}
so that in order to obtain $|V_{us}|$ we need to determine $f^+(0)$. In the last four years, following ref.~\cite{Becirevic:2004ya}, lattice QCD calculations of $f^+(0)$ have been undertaken using dynamical simulations with $N_f=2$ or $N_f=2+1$ flavours of sea quarks\cite{Boyle:2007qe, Okamoto:2004df, Hashimoto:2005am, Dawson:2006qc, Brommel:2007wn}, thus enabling the extraction of $V_{us}$\,.

In this paper we investigate the behaviour of $K_{\ell 3}$ decay amplitudes with the masses of the $u$ and $d$ quarks. This is an interesting problem in itself, but the immediate motivation is the need to extrapolate the lattice results for $f^+(0)$, obtained with $u$ and $d$ quarks heavier than the physical ones (typically with pions with masses in the range $m_\pi\gtrsim 300\mev$), to their physical values.

The mass dependence of the form factors is studied below using $\SU(2)_L\times \SU(2)_R$ chiral perturbation theory (ChPT) at next-to-leading order (NLO)~\footnote{For compactness of notation in the remainder of this paper we refer to $\SU(n)_L\times \SU(n)_R$ ChPT (for $n=2$ or 3) as $\SU(n)$ ChPT.}. Conventionally, following the seminal paper of Gasser and Leutwyler~\cite{Gasser:1984ux}, it has been $\SU(3)$ ChPT which has been applied to the study of $\kl3$ decay amplitudes. However, following the study of the quark mass dependence of physical quantities computed in a lattice simulation using Domain Wall Fermions~\cite{Allton:2008pn}, together with our colleagues from the RBC and UKQCD collaborations we concluded that it may be better to use $\SU(2)$ ChPT, at least for some quantities. This conclusion is primarily based on the large one-loop effects in $\SU(3)$ ChPT found for the leptonic decay constant of \lq pions\rq\ with masses in the range in which the simulations were performed. Note also that
the strange quark mass ($m_s$) in lattice simulations can be chosen to be at its physical value and so ChPT is not needed to perform the corresponding extrapolation (although, since the bare strange quark mass is chosen before the simulation is undertaken, in practice there may have to be a small extrapolation to correct for the difference between the $m_s$ used in the simulation and its physical value; $\SU(3)$ ChPT may provide useful guidance for this). Of course, in using $\SU(2)$ rather than $\SU(3)$ ChPT we sacrifice some symmetry and therefore some information.

In ref.~\cite{Allton:2008pn}, together with the RBC and UKQCD collaborations, we developed and used $\SU(2)$ ChPT for kaon physics; in particular we studied the dependence on the pion mass of the mass of the kaon $m_K$, the leptonic decay constant $f_K$ and the $B_K$-parameter which contains the non-perturbative QCD effects in $K^0$\,--\,$\bar{K}^0$ mixing. We recall the main features of the formalism in section~\ref{sec:formalism}. An important difference between $\SU(3)$ and $\SU(2)$ ChPT is that with $\SU(2)$ ChPT, powers of $\bar{m}_K^2/\lchi^2$, where $\bar{m}_K$ is the mass of the kaon in the limit $m_u=m_d=0$ and $\lchi$ is the scale of chiral symmetry breaking, are absorbed into the low-energy constants (LECs). In $\SU(3)$ ChPT at $n$-loop order on the other hand, there remain errors of $O((\mkbar/\lchi)^{2(n+1)})$. The corresponding uncertainties in $\SU(2)$ ChPT are of $O((m_\pi/\lchi)^{2(n+1)})$ and $O((m_\pi/\mkbar)^{2(n+1)})$.

In this paper we study two aspects of the chiral behaviour of the $\kl3$ form factors:
\begin{enumerate}
\item \textbf{The behaviour of $f^+(0)=f^0(0)$ with $m_u$ and $m_d$.} In order to determine $V_{us}$ from eq.~(\ref{eq:vusfplus}) we need the form factor at $q^2=0$ for physical values of the quark masses. The result presented in eq.~(\ref{eq:f0qsq0}) below represents the behaviour of the form factor with the pion mass at NLO (one-loop order) and can be used to extrapolate the lattice results obtained at larger values of $m_u=m_d$ to the physical
point.

In order to derive eq.~(\ref{eq:f0qsq0}) one has to overcome a subtlety. At $q^2=0$,
\[ 2p_\pi\cdot p_K=m_K^2+m_\pi^2 = \mkbar^2+O(m_\pi^2)\,,\]
so that $E_\pi$, the energy of the pion in the rest frame of the kaon,
is approximately equal to $m_K/2$ and is not small (i.e. it is not of
$O(m_\pi)$) for $m_K^2\gg m_\pi^2$. Since $\SU(2)$ ChPT is an expansion in powers of masses and momenta of the pions, the fact that the external pion in $K\to\pi$ semileptonic decays is \textit{hard} complicates this power counting. Nevertheless, by integrating by parts, we show in section\,\ref{sec:fqsq0} that an expansion in small masses and momenta of $O(m_\pi)$ is possible and results in eq.~(\ref{eq:f0qsq0}). This is possible because the chiral logarithms arise from soft regions of phase-space for the {\em internal} pions.
\begin{table}[t]
\begin{center}
\begin{tabular}{|l|l|l|}\hline
$m_\pi$ [MeV]&$\qsqmax$ [\,GeV$^2$]&$f^0(\qsqmax)$\\ \hline
\ 671(11)&0.00235(4)& 1.00029(6)\\
\ 556(9)&0.01252(20)&1.00192(34)\\
\ 416(7)&0.03524(62)&1.00887(89)\\
\ 329(5)&0.06070(107)&1.02143(132)\\ \hline
\end{tabular}
\caption{Results for $f^0(\qsqmax)$ from ref.~\cite{Boyle:2007qe} at four values of the quark masses, corresponding to the pion masses given in the first column. \label{tab:prlresults}}
\end{center}
\end{table}
\item \textbf{The behaviour of $f^0(\qsqmax)$ with $m_u$ and $m_d$.} The maximum physical value of $q^2$ is $(m_K-m_\pi)^2$, corresponding to the pion and kaon both at rest. Using the double ratio techniques proposed in ref.~\cite{Becirevic:2004ya}, $f^0(\qsqmax)$ is evaluated with remarkable precision in lattice simulations. For illustration we reproduce in table\,\ref{tab:prlresults} the results from the RBC and UKQCD collaborations' simulation on a $24^3$ spatial lattice\,\cite{Boyle:2007qe}. The point which we particularly wish to underline here is that in the $\SU(2)$ chiral limit ($m_u=m_d=0$), the Callan-Treiman relation \cite{Callan:1966hu} implies that
\begin{equation}\label{eq:callantreiman}
f^0(\qsqmax)\quad\underset{m_\pi^2\to 0}{\longrightarrow}\quad\frac{f^{(K)}}{f}\,,
\end{equation}
where $f^{(K)}$ and $f$ are the kaon and pion leptonic decay constants in the $\SU(2)$ chiral limit. The Callan-Treiman relation was derived for the unphysical value of $q^2=m_K^2-m_\pi^2$, nevertheless, in the chiral limit it also holds for $\qsqmax$. We shall show however, that the corrections to the relation are of $O(m_\pi)$ and not the standard ChPT corrections of $O(m_\pi^2)$. The ratio of physical decay constants is $f_K/f_\pi\simeq 1.2$ and in the chiral limit it is a little larger, e.g. the lattice study of ref.~\cite{Allton:2008pn} finds a ratio $f^{(K)}/f\simeq 1.26$~\footnote{Enno Scholz private communication. This particular result is not quoted directly in \cite{Allton:2008pn}.}. In ref.~\cite{Boyle:2007qe} the entries in table~\ref{tab:prlresults} were obtained with a strange quark mass which is a little larger than the physical one and the corresponding value of $f^{(K)}/f$ is about 1.28. This is the value which we use in the numerical estimates below, together with $f\simeq 115\mev$, which is the central value found in \cite{Allton:2008pn}. We restrict the comparison of ChPT with table~\ref{tab:prlresults} to the entries with $m_\pi=329$ and $429\mev$, since our experience from ref.~\cite{Allton:2008pn} is that one-loop ChPT is less reliable at the heavier masses.
The values of $f^0(\qsqmax)$ in table\,\ref{tab:prlresults} are equal to 1 within 2\% or so, and although they are increasing as the quark masses are reduced, the observed increase is very slow indeed. As $m_\pi$ decreases from $670$ to $330\mev$, $f^0(\qsqmax)$ increases only from 1.00 to 1.02 which is still a long way from the expected value of about 1.28 in the chiral limit. We investigate the chiral behaviour of $f^0(\qsqmax)$ up to one-loop order in section\,\ref{sec:fqsqmax} and find that the chiral logarithm has a large coefficient but the wrong sign to account for the extrapolation to $f^{(K)}/f$. The coefficient of the linear term in $m_\pi$, which is not calculable in $\SU(2)$ ChPT, can be estimated by \textit{converting} the $\SU(3)$ results of ref.~\cite{Gasser:1984ux} to the $\SU(2)$ theory. We find that it is large with the correct sign but predicts too large a ratio between $f^0(\qsqmax)$ in the $\SU(2)$ chiral limit and at the masses where lattice simulations are performed. We also study the full prediction from $\SU(3)$ ChPT, which reproduces qualitatively (and semi-quantitatively) the observed behaviour.
\end{enumerate}
The plan for the remainder of this paper is as follows. In the
following section we briefly recall some of the main features of
$\SU(2)$ ChPT for kaon physics. Within this context, we also derive
eq.~(\ref{eq:callantreiman}). Sections~\ref{sec:fqsq0} and
\ref{sec:fqsqmax} contain the studies of the chiral behaviour of the
form factors at $q^2=0$ and $\qsqmax$ respectively. Our calculations
have some overlap with those of semileptonic decays of
$B$-mesons~\cite{Wise:1992hn,Burdman:1992gh,Wolfenstein:1992xh,Burdman:1993es,Falk:1993fr,Becirevic:2002sc,Becirevic:2003ad}
and we discuss the similarities and differences in
section~\ref{sec:bcomparison}.

\section{$\SU(2)$ Chiral Perturbation Theory for Kaons}\label{sec:formalism}

We start by briefly summarising the formalism introduced in section
II.B of ref.~\cite{Allton:2008pn} which we apply in the following sections to $K_{\ell 3}$ decays. We write the pion matrix, the quark mass matrix and the kaon fields in the form:
\begin{equation}\label{eq:phidef}
\phi=\begin{pmatrix}\pi^0/\sqrt{2}&\pi^+\\\pi^-&-\pi^0\sqrt{2}\end{pmatrix},\quad M=\begin{pmatrix}m_l&0\\ 0&m_l\end{pmatrix}\quad\textrm{and}\quad K=\begin{pmatrix}K^+\\ K^0\end{pmatrix}\ .\end{equation}
We work in the isospin limit so that $m_l$ represents $m_u=m_d$. The pion matrices $\xi$ and $\Sigma$ are defined in the standard way:
\begin{equation}\label{eq:xisigmadef}
\xi=\exp(i\phi/f)\quad\textrm{and}\quad\Sigma=\xi^2\,,
\end{equation}
where $f$ is the pion decay constant in the $\SU(2)$ chiral limit, $m_u=m_d=0$. As with all LECs in $\SU(2)$ ChPT, $f$ depends on $m_s$, the mass of the strange quark. Throughout this paper we define the pion and kaon decay constants using a normalization in which the physical value for the pion is $f_\pi\simeq 131\mev$.

We need to construct the chiral Lagrangian and operators which transform in a specified way under $\SU(2)$ chiral transformations out of the fields in eqs.~(\ref{eq:phidef}) and (\ref{eq:xisigmadef}). Under global left and right handed transformations, $L$ and $R$ respectively, these fields transform as follows:
\begin{equation}\label{eq:transformations}
\xi\to L\xi U^\dagger = U\xi R^\dagger,\quad\Sigma\to L\Sigma R^\dagger\quad\textrm{and}\quad K\to UK\,,
\end{equation}
where $U$ is a function of $L,\,R$ and the meson fields which reduces to a global vector transformation when $L=R$. From the transformations in eq.~(\ref{eq:transformations}) we construct operators with the required flavour and chiral quantum numbers.

The pion Lagrangian at lowest order is well known:
\begin{equation}\label{eq:lpipi}
L_{\pi\pi}^{(2)}=\frac{f^2}{8}\,\textrm{tr}\,\{\partial_\mu\Sigma\partial^\mu\Sigma^\dagger\}+ \frac{f^2B}{4}\textrm{tr}\,\{M^\dagger\Sigma+M\Sigma^\dagger\}\,,
\end{equation}
where $B$ is the standard lowest order LEC and to this order $m_\pi^2=2Bm_l$. For the interactions of kaons, which are not considered soft in the $\SU(2)$ ChPT formalism, with soft pions the chiral Lagrangian has been introduced by Roessl~\cite{Roessl:1999iu}
and at lowest order is given by
\begin{equation}\label{eq:lpik}
L_{\pi K}^{(1)}=D_\mu K^\dagger\, D^\mu K - \bar{m}_K^2 K^\dagger K\,,
\end{equation}
where the covariant derivative $D_\mu$ is constructed using the vector field $V_\mu$,
\begin{equation}\label{eq:vdef}
V_\mu=\frac12\left(\xi^\dagger\partial_\mu\xi+\xi\partial_\mu\xi^\dagger\right)\to UV_\mu U^\dagger+U\partial_\mu U^\dagger,
\end{equation}
and is defined by
\begin{equation}\label{eq:Ddef}
D_\mu K=\partial_\mu K+V_\mu K\to UD_\mu K\,.
\end{equation}

In the following it will be necessary also to introduce the pion axial vector field defined by
\begin{equation}\label{eq:adef}
A_\mu=\frac{i}{2}\left(\xi^\dagger\partial_\mu\xi-\xi\partial_\mu\xi^\dagger\right)\to UA_\mu U^\dagger\,.
\end{equation}
When constructing Feynman diagrams from the $\pi K$ Lagrangian and the effective theory local operators, we expand the vector and axial fields in terms of the pion fields,
\begin{equation}\label{eq:va_expand}
V_\mu=\frac{1}{2f^2}\left[\phi,\partial_\mu\phi \right]+\cdots\quad\textrm{and}\quad A_\mu=-\frac{1}{f}\,\partial_\mu\phi+\cdots
\end{equation}
so that the first term in the expansion of the vector field contains two pions and that for the axial field starts with a single pion.

Similar calculations for $B\to\pi$ and $B\to K$ semileptonic decays
using heavy meson ChPT were undertaken in
ref.~\cite{Falk:1993fr,Becirevic:2002sc} and we will discuss the
similarities and differences with kaon decays in more detail in
section~\ref{sec:bcomparison}. Here we simply point out that in the
heavy meson ChPT, the limit $m_b\to\infty$ is taken before performing
the chiral expansion. The resulting spin symmetry implies that the
$B^\ast$ vector meson is degenerate with the pseudoscalar $B$, and so
the $B^\ast B\pi$ interactions (where the pion is soft), and hence
diagrams containing $B^\ast$ propagators, must also be included. In
our kaon case, the $K^\ast-K$ mass splitting is considered to be of
$O(\lqcd)$ and so the corresponding diagrams are absent.

\subsection{Lowest Order $\Delta S=1$ vector and axial currents}

We end this section by discussing the lowest order $\Delta S=1$ vector and axial currents in the effective theory. As already mentioned in the Introduction, it will not be enough to consider only the lowest order contributions and we will have to extend the present discussion in the following two sections.

The left handed QCD $\Delta S=1$ current is
\begin{equation}\label{eq:qcdldef}
J_\mu^L=\bar q_L\gamma_\mu s_L=\bar q\gamma_\mu\frac{(1-\gamma_5)}{2}s\,,
\end{equation}
where $q=u$ or $d$. It is convenient to promote $q$ to be a 2-component vector with components $u$ and $d$ and to introduce a 2-component constant spurion vector $h$ in order to be able to project $u$ and $d$ as required; specifically we write the left-handed current as
\begin{equation}\label{eq:qcdl_spurion}
\bar q\,h\ \gamma_\mu\frac{(1-\gamma_5)}{2}s\,.
\end{equation}
The current in eq.~(\ref{eq:qcdl_spurion}) would be invariant under $\SU(2)_L$ transformations if $h$ transformed as $h\to Lh$. We now construct the form of the left-handed current in the effective theory. This is a linear combination of all operators which are linear in $h$ and which would be invariant under $\SU(2)_L$ transformations if $h$ transformed as above. At lowest order in the chiral expansion we identify two possible independent terms and, following the notation of ref.~\cite{Allton:2008pn}, we write the left-handed current as
\begin{equation}\label{eq:la1la2}
J_\mu^L=-L_{A1}(D_\mu K)^\dagger\xi^\dagger h + iL_{A2}K^\dagger A_\mu\xi^\dagger h\,,
\end{equation}
where $L_{A1}$ and $L_{A2}$ are LECs and $A_\mu$ is the pion axial current defined in eq.~(\ref{eq:adef}). Note that since $A_\mu\xi^\dagger=-iD_\mu\xi^\dagger$ (and also $A_\mu\xi=iD_\mu\xi$), no new independent operator is obtained by replacing $A_\mu$ by the covariant derivative in the second term on the right-hand side of eq.~(\ref{eq:la1la2}).

For the right-handed current, we take the transformation on $h$ to be $h\to Rh$ and obtain two possible operators at lowest order,
\begin{equation}
(D_\mu K)^\dagger \xi h \quad\textrm{and}\quad K^\dagger A_\mu\xi h\,,
\end{equation}
which transform as $\frac12\bar q h\gamma_\mu(1+\gamma_5)\,s$. Noting that parity transformations, under which
\begin{equation}\label{eq:parity}
K\to -K,\quad \xi\to\xi^\dagger, \quad A_\mu\to -A_\mu,
\end{equation}
transform $J_\mu^L$ into $J_\mu^R$ (where $J_\mu^R$ is the right-handed current), so that the same LECs, $L_{A1}$ and $L_{A2}$ appear in the right-handed current,
\begin{equation}
J_\mu^R=L_{A1} (D_\mu K)^\dagger\xi h+iL_{A2}K^\dagger A_\mu\xi h\,.
\end{equation}
In some applications, the $L_{A2}$ term can be considered to be sub-leading since the derivative is on the pion rather than the kaon field. The vector ($J_\mu$) and axial-vector ($J_\mu^5$) currents can now readily be determined:
\begin{eqnarray}\label{eq:j}
J_\mu=J_\mu^R+J_\mu^L&=&L_{A1}(D_\mu K)^\dagger (\xi-\xi^\dagger)h + iL_{A2}\,K^\dagger A_\mu(\xi+\xi^\dagger)h\\
\label{eq:j5}
J_\mu^5=J_\mu^R-J_\mu^L&=&L_{A1}(D_\mu K)^\dagger (\xi+\xi^\dagger)h + iL_{A2}\,K^\dagger A_\mu(\xi-\xi^\dagger)h\,.
\end{eqnarray}
The LEC $L_{A1}$ appears in both the vector and axial-vector currents
and we will see in section~\ref{sec:fqsqmax} that it is this feature
which allows us (in the $\SU(2)$ chiral limit) to relate the
$K\to\mathrm{vacuum}$ matrix element of the axial-vector current and
the $K\to\pi$ matrix element of the vector current and hence to derive
eq.~(\ref{eq:callantreiman}). Evaluating the $K\to\mathrm{vacuum}$
matrix element in the chiral limit immediately shows us that
\begin{equation}\label{eq:fkla1}
f^{(K)}=2 L_{A1}.
\end{equation}

The symmetry arguments used here apply also to other flavours, so that eq.(\ref{eq:callantreiman}) can be generalised to $D$ and $B$ mesons and this is briefly discussed in section\,\ref{sec:bcomparison} below.

\section{$\kl3$ form factors at $q^2=0$}\label{sec:fqsq0}

As discussed in the Introduction, in order to study the chiral behaviour of the form factor $f^0$ at $q^2=0$ we have to deal with the fact that in this case $2p_K\cdot p_\pi\simeq m_K^2$ and hence we cannot neglect operators with an arbitrary numbers of derivatives on the external pion field. This situation is reminiscent of the light-cone dominated process of deep-inelastic scattering. To illustrate the point consider the matrix element
\begin{equation}\label{eq:od1def}
\langle\,\pi(p_\pi)\,|\,(D_\nu D_\mu K)^\dagger D^\nu (\xi-\xi^\dagger)h\,|\,\bar{K}(p_K)\,\rangle
\end{equation}
at $q^2=0$. In spite of the additional derivative acting on the pion field relative to the first term in $J_\mu$ in eq.~(\ref{eq:j}), the matrix element in (\ref{eq:od1def}) does give a leading contribution in the chiral expansion since by inspection we see that there is a contribution of $p_K\cdot p_\pi$ times the matrix element of $(D_\mu K)^\dagger (\xi-\xi^\dagger)h$. Nevertheless, as we now show, the leading contribution is simply proportional to the matrix element of $(D_\mu K)^\dagger (\xi-\xi^\dagger)h$ (with a constant of proportionality which depends on $m_s$ but not on $m_{u,d}$) and so the chiral logarithms are the same and the number of LECs remains the same. To see this, note that at $q^2=0$,
\begin{eqnarray}
0&=&\partial^2\langle\,\pi(p_\pi)\,|\,(D_\mu K)^\dagger (\xi-\xi^\dagger)h\,|\,\bar{K}(p_K)\,\rangle\nonumber\\
&=&\langle\,\pi(p_\pi)\,|\,(D^2D_\mu K)^\dagger(\xi-\xi^\dagger)h+2(D_\nu D_\mu K)^\dagger D^\nu(\xi-\xi^\dagger)h\nonumber\\
&&+(D_\mu K)^\dagger D^2(\xi-\xi^\dagger)h\,|\,\bar{K}(p_K)\,\rangle
\end{eqnarray}
so that
\begin{eqnarray}\label{eq:qsq0_aux1}
\lefteqn{\langle\,\pi(p_\pi)\,|\,(D_\nu D_\mu K)^\dagger D^\nu
  (\xi-\xi^\dagger)h\,|\,\bar{K}(p_K)\,\rangle =}\\
&&-\frac12\left\{
\langle\,\pi(p_\pi)\,|\,(D^2 D_\mu K)^\dagger (\xi-\xi^\dagger)h\,|\,\bar{K}(p_K)\,\rangle+
\langle\,\pi(p_\pi)\,|\,(D_\mu K)^\dagger D^2(\xi-\xi^\dagger)h\,|\,\bar{K}(p_K)\,\rangle
\right\}\,.\nonumber
\end{eqnarray}
Before discussing the chiral behaviour of the operators on the right-hand side of eq.~(\ref{eq:qsq0_aux1}) we clarify our power counting. The external pion is \textit{hard} in the sense that $2p_\pi\cdot p_K\simeq m_K^2$ and so we need to keep $p_\pi\cdot p_K$ to any power. This is the reason the matrix element in eq.~(\ref{eq:od1def}) is of leading order. We accept that the corrections of $O(m_\pi^2)$ are multiplied by an unknown constant and so we do not attempt to calculate such terms. We do however, calculate the chiral logarithms, i.e. the corrections of $O(m_\pi^2\log(m_\pi^2))$ and in order to evaluate these we can treat the internal pion momenta as being \textit{soft}, i.e. of $O(m_\pi)$.

The operator in the second term on the right hand side of
eq.~(\ref{eq:qsq0_aux1}) contains the insertion
$D^2(\xi-\xi^\dagger)$\,. This leads to a contribution which is
suppressed by a factor of $m_\pi^2$, with no chiral logarithm
proportional to $m_\pi^2\log(m_\pi^2)$. Thus, up to the order to which we are working, we only need to consider the first term on the right-hand side of eq.~(\ref{eq:qsq0_aux1}), where we can replace $(D^2 D_\mu K)^\dagger$ by $(D_\mu D^2 K)^\dagger=-\bar{m}_K^2\, (D_\mu K)^\dagger$ up to terms which are suppressed by $m_\pi^2$. Note that the commutator $[D_\mu,D_\nu]$ contains two derivatives acting on two different pion fields, at least one of which must be Wick contracted to give a soft internal propagator. This leads to a suppression of $O(m_\pi^2)$ without chiral logarithms and we arrive at the useful result that, up to corrections of $O(m_\pi^2)$ (without chiral logarithms):
\begin{equation}
\langle\,\pi(p_\pi)\,|\,(D_\nu D_\mu K)^\dagger D^\nu (\xi-\xi^\dagger)h\,|\,\bar{K}(p_K)\,\rangle=
-\frac{\bar{m}_K^2}{2}\,
\langle\,\pi(p_\pi)\,|\,(D_\mu K)^\dagger (\xi-\xi^\dagger)h\,|\,\bar{K}(p_K)\,\rangle\,.
\end{equation}
Thus in order to include the contribution of the matrix element (\ref{eq:od1def}) to the $K\to\pi$ form factor $f^0(0)$, including the one-loop chiral logarithms, it is sufficient to replace the LEC $L_{A1}$ in the definition of the vector current in eq.~(\ref{eq:j}) by an unknown coefficient which depends on $m_s$ but not on the light-quark masses.

The discussion of the matrix element in (\ref{eq:od1def}) presented explicitly above can be generalised to other operators. Leading-order operators can have any number of covariant derivatives on the external pion field. If the Lorentz index of a covariant derivative acting on the external pion field is contracted with another derivative on the external pion then we obtain a non-leading correction of $O(m_\pi^2)$. Similarly if it is contracted with a derivative on a pion in an internal loop we also obtain a similar suppression. Finally if it is contracted with a derivative on the kaon field then we can reduce it to an operator which is proportional to the leading operator by integrating by parts as above. Note also that the kaon mass-squared, $m_K^2$, has no chiral logarithms of the form $m_\pi^2\log(m_\pi^2)$ and so no chiral logarithms are introduced by using the equations of motion.

From this discussion we see that to leading order at $q^2=0$ we have 
\begin{eqnarray}\label{eq:tildelas}
\lefteqn{\langle\,\pi(p_\pi)\,|\,\bar{q}\,\gamma_\mu(1-\gamma_5)s\,|\,\bar{K}(p_K)\,\rangle =}\\
&&\langle\,\pi(p_\pi)\,|\,\tilde{L}_{A1}(D_\mu K)^\dagger (\xi-\xi^\dagger)h+\tilde{L}_{A2}
K^\dagger A_\mu(\xi+\xi^\dagger)h\,|\,\bar{K}(p_K)\,\rangle\nonumber
\end{eqnarray}
where we recall that $\tilde{L}_{A1}$ and $\tilde{L}_{A2}$ are unknown constants which cannot be obtained from  $L_{A1}$ and $L_{A2}$ alone. They depend on $m_s$ but not on the light-quark masses and hence we treat them as LECs, noting however that they are only relevant for the case $q^2=0$. As a result of the fact that the matrix element at $q^2=0$ is written in terms of $\tilde{L}_{A1}$ and $\tilde{L}_{A2}$ rather than $L_{A1}$ and $L_{A2}$, we lose the connection to $f^{(K)}/f$ in this case.

\subsection{The Chiral Logarithms}

\begin{figure}
\begin{center}
\begin{picture}(410,100)(0,0)
\ArrowLine(0,50)(30,50)\ArrowLine(30,50)(60,50)\GCirc(30,50){4}{0.8}
\Text(15,44)[t]{{\small$p_K$}}\Text(45,44)[t]{{\small$p_\pi$}}
\Text(30,25)[t]{{\small(a)}}
\ArrowLine(110,50)(140,50)\ArrowLine(140,50)(170,50)
\Oval(140,73)(20,8)(0)\GBox(136,46)(144,54){0.8}\GCirc(140,93){4}{0.8}
\Text(125,44)[t]{{\small$p_K$}}\Text(155,44)[t]{{\small$p_\pi$}}
\Text(140,25)[t]{{\small(b)}}\Text(128,73)[r]{{\scriptsize K}}
\Text(152,73)[l]{{\small $\pi$}}
\ArrowLine(220,50)(250,50)\ArrowLine(250,50)(280,50)
\Oval(250,64)(12,8)(0)\GCirc(250,50){4}{0.8}
\Text(235,44)[t]{{\small$p_K$}}\Text(265,44)[t]{{\small$p_\pi$}}
\Text(250,25)[t]{{\small(c)}}\Text(250,81)[b]{{\small$\pi$}}
\ArrowLine(330,50)(360,50)\Line(360,50)(380,50)
\ArrowLine(380,50)(410,50)\GCirc(360,50){4}{0.8}
\Oval(380,65)(12,6)(0)\GBox(376,46)(384,54){0.8}\Text(380,81)[b]{{\small$\pi$}}
\Text(345,44)[t]{{\small$p_K$}}\Text(395,44)[t]{{\small$p_\pi$}}
\Text(380,25)[t]{{\small(d)}}
\end{picture}
\caption{Diagrams contributing to the $K\to\pi$ matrix elements at tree level (diagram (a)) and at one-loop level (diagrams (b), (c) and (d)). The grey circle represents the insertion of the $K\to\pi$ vector current and the grey box the insertion of the $KK\pi\pi$ vertex (diagram (b)) or the four-pion vertex (diagram (d)) from the strong Lagrangian.}\label{fig:diags}
\end{center}
\end{figure}
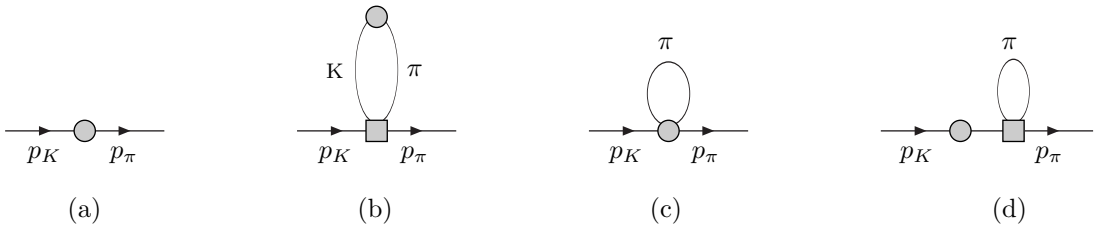

The tree level contribution to the matrix element in eq.~(\ref{eq:tildelas}) is
\begin{equation}\label{eq:tildelastree}
\langle\,\pi(p_\pi)\,|\,\bar{q}\,\gamma_\mu(1-\gamma_5)s\,|\,\bar{K}(p_K)\,\rangle = \frac{2\tilde{L}_{A1}}{f}\,p_{K\,\mu}
+\frac{2\tilde{L}_{A2}}{f}\,p_{\pi\,\mu}\,.
\end{equation}
This contribution is represented diagrammatically in fig.~\ref{fig:diags}(a).

In order to obtain the chiral logarithms at one-loop order we need to evaluate the diagrams in fig.~\ref{fig:diags}(b), (c) and (d), where fig.~\ref{fig:diags}(d) represents the contribution to the pion's wave-function renormalization. There is no one-loop chiral logarithm contributing to the kaon's wave function renormalization which can be deduced from the structure of the $D_\mu KD^\mu K$ term in the strong Lagrangian $L_{\pi K}^{(1)}$ in (\ref{eq:lpik}). The $KK\pi\pi$ vertex arises when one keeps a partial derivative $\partial_\mu$ from one of the $D_\mu$ factors and the current $V_\mu$ from the other. From eq.~(\ref{eq:va_expand}) we see that the expansion of $V_\mu$ starts with two pion fields, on one of which there is a  single derivative. This derivative corresponds to a single momentum in the numerator of the tadpole loop and hence the momentum integration is odd and gives zero. The chiral logarithms from each of the diagrams in fig.~\ref{fig:diags} are presented in table\,\ref{tab:logs_qsq0} together with the total.

Gasser and Leutwyler have calculated the chiral logarithms in the $\SU(3)$ theory as a function of $q^2$~\cite{Gasser:1984ux}. In this case the power counting is different from that here, in that $m_K$ is also considered to be small. It is instructive to check our calculation by \textit{converting} the $\SU(3)$ results to $\SU(2)$, using eq.~(2.6) of ref.~\cite{Gasser:1984ux} and the expression for $\bar J$ in eq.~(A.7) of ref.~\cite{Gasser:1984gg}. Expanding the Gasser-Leutwyler results in powers of $m_\pi^2$, we confirm that the total one-loop chiral logarithms in table~\ref{tab:logs_qsq0} are indeed correct.

\begin{table}[t]
\begin{center}
\begin{tabular}{|c|c|}\hline
Diagram & Result\\ \hline
\rule[-3mm]{0mm}{9mm}fig.~\ref{fig:diags}(a) & $\frac{2\tilde{L}_{A1}}{f}\,p_{K\mu}+\frac{2\tilde{L}_{A2}}{f}\,p_{\pi\mu}$\\
\rule[-3mm]{0mm}{9mm}fig.~\ref{fig:diags}(b) & $-\frac{2\tilde{L}_{A1}}{f}\,(p_{K\mu}-p_{\pi\mu})\,L$\\
\rule[-3mm]{0mm}{9mm}fig.~\ref{fig:diags}(c) & $\frac{2\tilde{L}_{A1}}{f}\,\left(-\frac{5}{12}p_{K\mu}-p_{\pi\mu}\right)+\frac{2\tilde{L}_{A2}}{f}\, \left(-\frac{17}{12}p_{\pi\mu}\right)L$\\
\rule[-3mm]{0mm}{9mm}fig.~\ref{fig:diags}(d) &$\frac23\left(\frac{2\tilde{L}_{A1}}{f} p_{K\mu}+\frac{2\tilde{L}_{A2}}{f}p_{\pi\mu}\right)L$\\
\hline
\rule[-3mm]{0mm}{9mm}TOTAL & $\left(\frac{2\tilde{L}_{A1}}{f}p_{K\mu}+\frac{2\tilde{L}_{A2}}{f}p_{\pi\mu}\right)
\,(1-\frac34L)$\\ \hline
\end{tabular}
\caption{\label{tab:logs_qsq0} Tree level expression and the one-loop chiral logarithms for the $K\to\pi$ matrix element at $q^2=0$.}
\end{center}
\end{table}

From table\,\ref{tab:logs_qsq0} we now have all the ingredients to write down the NLO expression for the $K\to\pi$ matrix element at $q^2=0$. The expression is
\begin{equation}\label{eq:meqsq0}
\langle\,\pi(p_\pi)\,|\,\bar{q}\gamma^\mu s\,|\,\bar{K}(p_K)\,\rangle=F_Kp_K^\mu\,\left[1-\frac34\,L+c_K\,m_\pi^2\right]+
F_\pi p_\pi^\mu\,\left[1-\frac34\,L+c_\pi\,m_\pi^2\right]\,,
\end{equation}
where $F_K=2\tilde{L}_{A1}/f$ and $F_\pi=2\tilde{L}_{A2}/f$ and $c_{K,\pi}$ are LECs. ($F_K$ and $F_\pi$ should not be confused with the leptonic decay constants for which we use the notation $f_K$ and $f_\pi$.) The chiral logarithm $L$ is defined by
\begin{equation}
L=\frac{m_\pi^2}{16\pi^2f^2}\,\log\left(\frac{m_\pi^2}{\mu^2}\right)\,,
\end{equation}
and the dependence of $L$ on $\mu$ is cancelled in expressions for physical quantities by the $\mu$-dependence of the LECs (e.g. in eq.~(\ref{eq:meqsq0}) the $\mu$ dependence of  $L$ is cancelled by that of $c_{\pi}$ and $c_K$).
Eq.~(\ref{eq:meqsq0}) implies that the chiral behaviour of the form factors is given by
\begin{eqnarray}
f^0(0)=f^+(0)&=&F_+\,(1-\frac34L+c_+m_\pi^2)\label{eq:f0qsq0}\\
f^-(0)&=&F_-\,(1-\frac34L+c_-m_\pi^2)\label{eq:fminusqsq0}
\end{eqnarray}
where again $F_\pm$ and $c_\pm$ are LECs, given in terms of the parameters present in eq.~(\ref{eq:meqsq0}) (for example, $F_\pm=\frac12\,(F_K\pm F_\pi)$\,).

Eq.~(\ref{eq:f0qsq0}) is the NLO $\SU(2)$ ChPT formula for extrapolating the lattice results for $f^0(0)=f^+(0)$ which are obtained at unphysical values of the up and down quark masses to the physical point. The two LECs $F_+$ and $c_+$ need to be determined by fitting the mass dependence of the measured values of $f^0(0)$ to (\ref{eq:f0qsq0}); the physical result of $f^0(0)$ is then readily obtained. Of course, using $\SU(3)$ ChPT the Ademollo-Gatto theorem \cite{Ademollo:1964sr} ensures that there are no LECs at one-loop order so that $F_+$ and $c_+$ are known and we can rewrite eq.~(\ref{eq:f0qsq0}) as
\begin{eqnarray}
f^0(0)=f^+(0)&=&\left(1-\frac{\bar{m}_K^2}{64\pi^2
  f^2}\left[5-12\log\frac43\right]\right)\times\nonumber\\
 &&\left(1+\frac{m_\pi^2}{64\pi^2 f^2}\left[-3\log\frac{m_\pi^2}{\mu^2}-4+9\log\frac43+3\log\frac{\bar{m}_K^2}{\mu^2}\right]\right)\,.
\label{eq:ademollogatto}\end{eqnarray}
The expressions for $F_+$ and $c_+$ in Eq.~(\ref{eq:ademollogatto}) are valid only at linear order in $m_s$; the numerical results for $f^+(0)$ at small pion masses were found to lie below the one-loop $\SU(3)$ ChPT expression~\cite{Boyle:2007qe}.

It is conventional for experimental results to be presented in terms of $|V_{us}|f^+(0)$ and so we have concentrated above on the chiral behaviour of the form factors at $q^2=0$. We can perform a similar analysis for any value of $q^2$ with $p_\pi\cdot p_K=O(m_K^2)$, but the effective LECs, i.e. the $F_\pm$ and $c_\pm$ depend on $q^2$.

\section{$\kl3$ form factor at $q^2=q^2_{\textrm{max}}$}\label{sec:fqsqmax}

We now turn our attention to the form factor $f^0(q^2_{\textrm{max}})$, where $q^2_{\textrm{max}}=(m_K-m_\pi)^2$. The tree-level diagram for the $K\to\pi$ decay is drawn in fig.\,\ref{fig:diags}(a) and its contribution to the amplitude is given in the first row of table\,\ref{tab:logs_qsqmax}. By setting $\mu=4$ for example and recalling that $2L_{A1}=f^{(K)}$ (see eq.\,(\ref{eq:fkla1})) we see that in the chiral limit $f^0(\qsqmax)=f^{(K)}/f$ and hence establish (\ref{eq:callantreiman}).

In the remainder of this section we try to understand why the lattice results for $f^0(q^2_{\textrm{max}})$ in table\,\ref{tab:prlresults} are significantly different from the value in the $\SU(2)$ chiral limit, $f^{(K)}/f$, and seem to be approaching this value very slowly, if at all. At $q^2_{\textrm{max}}$ the momentum of the external pion is small ($p_K\cdot p_\pi=m_Km_\pi$) and so the counting of contributions in terms of powers of $m_\pi$ is simpler than at $q^2=0$. However, close to the $\SU(2)$ chiral limit ($m_\pi=0$) the corrections to the $K\to\pi$ matrix element of the vector current are linear in $m_\pi$ (and not quadratic). To see this, consider for example, the matrix element in eq.~(\ref{eq:od1def}), $\langle\,\pi(p_\pi)\,|\,(D_\nu D_\mu K)^\dagger D^\nu (\xi-\xi^\dagger)h\,|\,\bar{K}(p_K)\,\rangle$, which is now manifestly linear in $m_\pi$. The coefficient of the linear term is not calculable directly in $\SU(2)$ ChPT.

Below we study the chiral behaviour of $f^0(q^2_{\textrm{max}})$ in three stages as follows:
\begin{enumerate}
\item[i)] We start in section\,\ref{subsec:chirallogs} by calculating the one-loop chiral logarithms, i.e. the corrections of $O(m_\pi^2\,\log(m_\pi)^2)$. This can be done within $\SU(2)$ ChPT.
\item[ii)] In order to estimate the remaining terms we use $\SU(3)$ ChPT. In the second stage, in section\,\ref{subsec:linear} below we estimate the
the coefficient of the linear term in $m_\pi$ by \textit{converting} the one-loop $\SU(3)$ expressions to $\SU(2)$.
\item[iii)] Finally, in section\,\ref{subsec:su3} we also estimate the quadratic terms using $\SU(3)$ ChPT .
\end{enumerate}
We will find that at $q^2_{\textrm{max}}$ the chiral corrections are very large and provide only a qualitative or perhaps a semi-quantitative, explanation of the observed chiral behaviour. Nevertheless, the calculations confirm that the differences of the lattice results from $f^{(K)}/f$ are reasonable.

\subsection{The chiral logarithms}\label{subsec:chirallogs}
\begin{table}[t]
\begin{center}
\begin{tabular}{|c|c|}\hline
Diagram & Result\\ \hline
\rule[-3mm]{0mm}{9mm}fig.~\ref{fig:diags}(a) & $\frac{2{L}_{A1}}{f}\,p_{K\mu}+\frac{2{L}_{A2}}{f}\,p_{\pi\mu}$\\
\rule[-3mm]{0mm}{9mm}fig.~\ref{fig:diags}(b) & $-\frac{2{L}_{A1}}{f}\,3(p_{K\mu}-p_{\pi\mu})\,L$\\
\rule[-3mm]{0mm}{9mm}fig.~\ref{fig:diags}(c) & $\frac{2{L}_{A1}}{f}\,\left(-\frac{5}{12}p_{K\mu}-p_{\pi\mu}\right)+\frac{2{L}_{A2}}{f}\, \left(-\frac{17}{12}p_{\pi\mu}\right)L$\\
\rule[-3mm]{0mm}{9mm}fig.~\ref{fig:diags}(d) &$\frac23\left(\frac{2{L}_{A1}}{f} p_{K\mu}+\frac{2{L}_{A2}}{f}p_{\pi\mu}\right)L$\\
\hline
\rule[-3mm]{0mm}{9mm}TOTAL & $\frac{2{L}_{A1}}{f}p_{K\mu}\left(1-\frac{11}{4}L\right)
+\frac{2{L}_{A2}}{f}p_{\pi\mu}\left(1-\frac{3}{4}L\right)
+\frac{2{L}_{A1}}{f}p_{\pi\mu}2L$\\ \hline
\end{tabular}
\caption{\label{tab:logs_qsqmax} Tree level expression and the one-loop chiral logarithms for the $K\to\pi$ matrix element at $q^2_{\textrm{max}}$.}
\end{center}
\end{table}
The chiral logarithms from each of the diagrams in
fig.~\ref{fig:diags} are presented in table\,\ref{tab:logs_qsqmax}.
From the table, choosing the Lorentz index $\mu=4$ and neglecting terms of $O(m_\pi^3)$, with or without logarithms, we deduce that the chiral behaviour of the form factor is of the form:
\begin{equation}\label{eq:f0qsqmaxchpt1}
f^0(q^2_{\textrm{max}})=\frac{f^{(K)}}{f}\left[1-\frac{11}{4}\,L+\frac{\lambda_1}{4\pi f}m_\pi+\frac{\lambda_2}{(4\pi f)^2}m_\pi^2+\cdots\right]
\end{equation}
where $\lambda_{1,2}$ are low energy constants which depend on the strange quark mass but not on the light quark masses. Again one can readily verify that the coefficient of the chiral logarithm in (\ref{eq:f0qsqmaxchpt1}) is indeed the result obtained by converting the general $\SU(3)$ formulae of Gasser and Leutwyler~\cite{Gasser:1984gg,Gasser:1984ux} to $\SU(2)$.

\begin{figure}[t]
\begin{center}
\includegraphics[width=0.5\textwidth,bb=7 0 235 165]{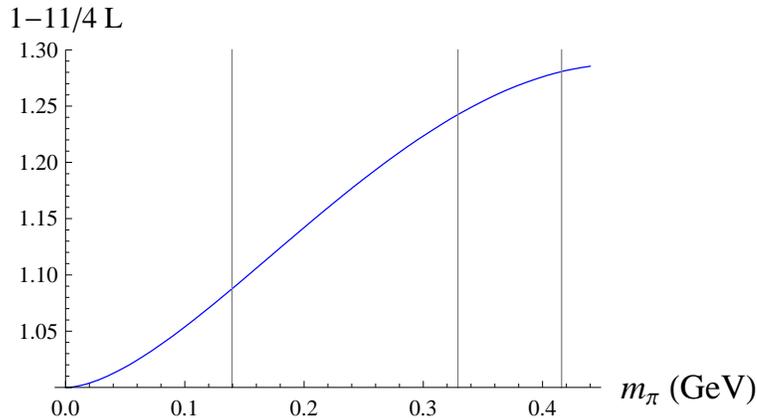}
\caption{Sketch of $1-11/4\,L$ as a function of the mass of the pion. The three vertical lines correspond (from left to right) to the physical pion mass and to the lightest two masses in the simulation of ref.~\cite{Boyle:2007qe}, $329$ and $416\mev$ respectively. $\mu$ was chosen to be $m_\rho=0.77\gev$. \label{fig:logsonly}}
\end{center}
\end{figure}

The coefficient $-11/4$ is large (for example, at $q^2=0$ in eq.~(\ref{eq:meqsq0}) the coefficient of $L$ is $-3/4$) and the term with the chiral logarithm does give a sizeable contribution in the region of pion masses between the physical one and that where the lattice simulations of ref.~\cite{Boyle:2007qe} were performed. In fig.~\ref{fig:logsonly} we sketch $1-11/4\,L$ with the physical mass of the $\rho$-meson as the scale $\mu$ and with $f=115\mev$ which is the central value found in \cite{Allton:2008pn}\,. The sign of the chiral logarithm however, is such as to make the form factor decrease as the mass of the pion is decreased towards the chiral limit, which is the opposite of what is required to account for the difference between the measured values of $f^0(q^2_{\textrm{max}})$ in table\,\ref{tab:prlresults} and $f^{(K)}/f$. Thus the chiral logarithms approximately double the size of the effect which should be explained.

\subsection{Linear term in $m_\pi$}\label{subsec:linear}
\begin{figure}[t]
\begin{center}
\includegraphics[width=0.45\textwidth,bb=50 0 239 141]{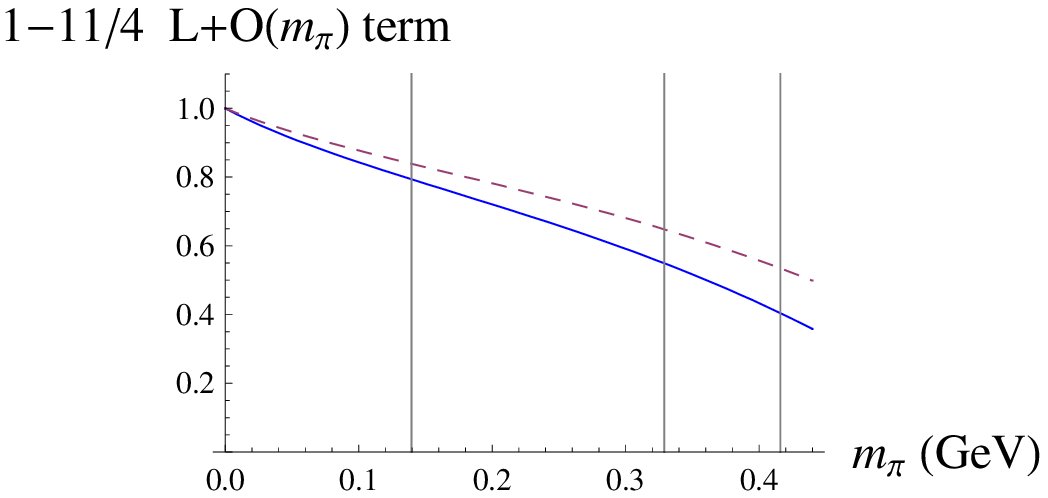}
\caption{Sketch of the expression in parentheses in eq.~(\ref{eq:f0qsqmaxchpt2}) with $\lambda_2=0$ as a function of the mass of the pion (solid curve). The three vertical lines correspond (from left to right) to the physical pion mass and to the lightest two masses in the simulation of ref.~\cite{Boyle:2007qe}, $329$ and $416\mev$ respectively. $\mu$ was chosen to be $m_\rho=0.77\gev$ and $f=115\mev$. The dashed line represents the expression in eq.~(\ref{eq:f0qsqmaxchpt3}). \label{fig:logsplusm}}
\end{center}
\end{figure}
We cannot evaluate $\lambda_1$ using $\SU(2)$ ChPT alone. To estimate whether the linear term in $m_\pi$ in (\ref{eq:f0qsqmaxchpt1}) can account for the difference of the measured form factors from  $f^{(K)}/f$ we convert the $\SU(3)$ results of Gasser and Leutwyler~\cite{Gasser:1984gg,Gasser:1984ux} to $\SU(2)$ ChPT. In this way we can obtain an approximate value of $\lambda_1$, using which we rewrite (\ref{eq:f0qsqmaxchpt1}) as:
\begin{eqnarray}
f^0(q^2_{\textrm{max}})&=&\frac{f^{(K)}}{f}\Bigg[1-\frac{11}{4}\,L-\frac{\bar{m}_Km_\pi}{(4\pi f)^2}\left(\frac{14}{3}+\frac{20}{9}\log\frac43-\frac89\sqrt{2}\arctan\sqrt{2}\right)\nonumber\\
&&\phantom{\frac{f^{(K)}}f\Bigg[}-\Big(\frac{f^{(K)}}{f}-1\Big)\frac{2m_\pi}{\bar{m}_K}+\frac{\lambda_2}{(4\pi f)^2}m_\pi^2+\cdots\Bigg]\,,
\label{eq:f0qsqmaxchpt2}\end{eqnarray}
where $\bar{m}_K$ is the mass of the kaon in the $\SU(2)$ chiral limit. Eq.~(\ref{eq:f0qsqmaxchpt2}) represents an approximation for $\lambda_1$ since, within $\SU(2)$ ChPT, $\lambda_1$ contains higher powers of $m_s$, whereas in (\ref{eq:f0qsqmaxchpt2}) we have kept only those from one-loop $\SU(3)$ ChPT. Setting $\lambda_2=0$ and neglecting higher order terms, we plot the expression in square parentheses in (\ref{eq:f0qsqmaxchpt2}) as a function of the pion mass as the solid curve in fig.~\ref{fig:logsplusm}, where we have set $f^{(K)}/f=1.28$, $f=115\mev$ and $\mu=m_\rho$. We notice that the linear term in $m_\pi$ does indeed change the sign, the value of the form-factor does increase as we approach the chiral limit. The effect is too large however, and since the $O(m_\pi)$ term is as large as 50-80\% in the region where we have data, the stability of the chiral expansion is likely to be questionable.

We write the converted expression from $\SU(3)$ ChPT as eq.~(\ref{eq:f0qsqmaxchpt2}) because this is the natural form for $\SU(2)$ ChPT. To illustrate that the result may depend significantly on the higher order terms we also plot, as the dashed curve in fig.~\ref{fig:logsplusm}, the expression
\begin{equation}\label{eq:f0qsqmaxchpt3}
1+\frac{f}{f^{(K)}}\left[-\frac{11}{4}\,L-\frac{\bar{m}_Km_\pi}{(4\pi f)^2}\left(\frac{14}{3}+\frac{20}{9}\log\frac43-\frac89\sqrt{2}\arctan\sqrt{2}\right)
-\Big(\frac{f^{(K)}}{f}-1\Big)\frac{2m_\pi}{\bar{m}_K}\right]
\end{equation}
which is equivalent to that in square parentheses (with $\lambda_2=0$)
in eq.~(\ref{eq:f0qsqmaxchpt2}) at one-loop order in $\SU(3)$ ChPT but differs by terms which are powers of $\bar{m}_K/(4\pi f)$. We make this choice because it is the form obtained directly from one-loop $\SU(3)$ ChPT. The difference in the curves in the region where we have lattice data is about 25-30\%, confirming that the uncertainties due to higher order terms are indeed likely to be large.

\subsection{$\SU(3)$ ChPT}\label{subsec:su3}

\begin{figure}[t]
\begin{center}
\includegraphics[width=0.5\textwidth,bb=12 0 279 193]{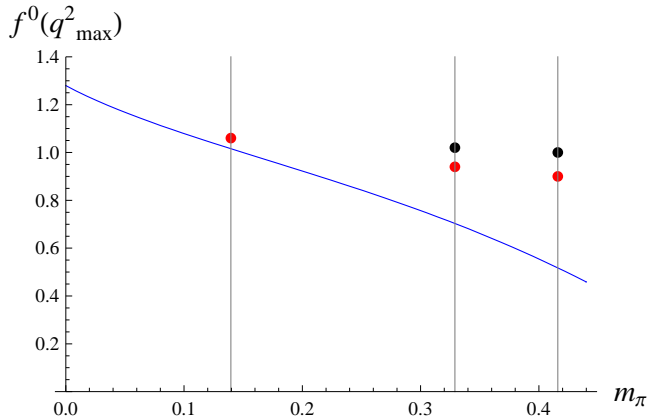}
\caption{The curve is a sketch of $f^0(q^2_{\textrm{max}})$ from eq.~(\ref{eq:f0qsqmaxchpt2}) with $\lambda_2=0$ as a function of the mass of the pion. The three vertical lines correspond (from left to right) to the physical pion mass and to the lightest two masses in the simulation of ref.~\cite{Boyle:2007qe}, $329$ and $416\mev$ respectively. For the curve $\mu$ was chosen to be $m_\rho=0.77\gev$ and $f=115\mev$. The black points are the lattice values from table~\ref{tab:prlresults} and the red points were obtained using $\SU(3)$ ChPT as described in the text.\label{fig:su3}}
\end{center}
\end{figure}

Finally we use the Gasser-Leutwyler $\SU(3)$ ChPT results to estimate the effect of the chiral extrapolation of $f^{(0)}$ from the quark masses used in the simulation in ref.~\cite{Allton:2008pn}. In this case, at one-loop order, there is one LEC, the Gasser-Leutwyler coefficient $L_5^r$, which is also the LEC which governs the $\SU(3)$ chiral behaviour of the ratio $f_K/f_\pi$; we can therefore use knowledge of the mass dependence of $f_K/f_\pi$ (e.g. from lattice simulations) to evaluate the term proportional to $L_5^r$. There is then no dependence on the scale $\mu$. We use eq.~(2.6) of ref.~\cite{Gasser:1984ux} to estimate the one-loop effects, but since the coefficients are large, the results depend on the precise procedure employed and on the choice of parameters (e.g. the physical value of $f_\pi$ or that in the $\SU(2)$ or $\SU(3)$ chiral limits), even though the differences are formally of higher order. As in section\,\ref{subsec:linear}, we find that one-loop $\SU(3)$ ChPT predicts that the mass dependence of $f^0(q^2_{\textrm{max}})$ is steeper than that expected from table\,\ref{tab:prlresults}; the dependence on the pion mass is however, much less steep than in section\,\ref{subsec:linear}. For example, using the physical value of the decay constant $f_\pi$ as the expansion parameter in the chiral expansion and taking the ratio of physical decay constants to be $f_K/f_\pi\simeq 1.20$, we find that the one-loop prediction for the form factor at the physical quark masses is about 1.06 (in the limit $m_\pi=0$ the result, of course, is $f^{(K)}/f\simeq 1.28$). Using the measured values of masses and decay constants from ref.~\cite{Allton:2008pn}, we find that, based on one-loop $\SU(3)$ ChPT, we would expect $f^0(q^2_{\textrm{max}})$ to be about 0.94 for $m_\pi=329\mev$ and 0.90 for $m_\pi=416\mev$ as compared to 1.02 and 1.01 in table~\ref{tab:prlresults}. To illustrate the flattening we plot in fig.~\ref{fig:su3}, the expected chiral behaviour of $f^0(q^2_{\textrm{max}})$ using the chiral logs and linear term (as in fig.~\ref{fig:logsplusm} but now multiplied by $f^{(K)}/f$). This is the curve in fig.~\ref{fig:su3}. We also exhibit the predicted values from $\SU(3)$ ChPT as calculated above (the 3 red points) and the two lattice points from table~\ref{tab:prlresults}.

Thus one-loop $\SU(3)$ ChPT, with the procedure we have employed, can provide a semi-quantitative explanation of the chiral behaviour observed in lattice simulations. Given the large effects we are finding at one-loop order, this is satisfying. We do stress however, that because of the large one-loop effects, the predictions are not very stable against varying the inputs into the chiral predictions, e.g. whether one uses the computed values of the decay constants (as we did above) or the values in the chiral limit. We have seen however, that the differences of the values in table~\ref{tab:prlresults} from $f^{(K)}/f$ are not unreasonable.

\section{Comparison with semileptonic B and D decays}\label{sec:bcomparison}

The chiral behaviour of semileptonic $B\to\pi$ decay amplitudes near
$q^2_\textrm{max}$ was studied in
refs.~\cite{Wise:1992hn,Burdman:1992gh,Wolfenstein:1992xh,Burdman:1993es}.
Chiral loop corrections were evaluated in standard chiral perturbation
theory in~\cite{Falk:1993fr} and extended to quenched and
partially-quenched cases in~\cite{Becirevic:2002sc,Becirevic:2003ad}.
For $B$-decays there is an additional scale, $m_b$, the mass of the
$b$-quark which is taken to be much larger than the typical hadronic
scale $\lqcd$. Indeed the calculations in
refs.~\cite{Falk:1993fr,Becirevic:2002sc,Becirevic:2003ad} are
performed by first taking the limit $m_b\to\infty$ (i.e.\ treating the
$b$-quark as being static) and then considering the chiral behaviour.
In that case additional diagrams to those in fig.~\ref{fig:diags} have
to be evaluated; in particular diagrams with $B^\ast$ propagators lead
to one-loop contributions with chiral logarithms. This is because in the static limit the $B$ and $B^\ast$ mesons are degenerate and so an on-shell $B$-meson can emit a soft pion and the resulting $B^\ast$-meson is also close to its mass-shell. The $K$ and $K^\ast$ mesons on the other hand, are not degenerate and so the corresponding contributions are already contained in the diagrams of fig.~\ref{fig:diags} and the
LECs. Thus the approach to the chiral limit depends on the order in which one takes the limits $m_b\to\infty$ and $m_\pi\to 0$ and we return to this point below.

The symmetry arguments used in section~\ref{sec:fqsqmax} which equate the semileptonic form factor $f^0(\qsqmax)$ in the chiral limit to $f^{(K)}/f$ can be generalised to other flavours and in particular to semileptonic $D$ and $B$-decays, so that,
\begin{equation}\label{eq:callantreimanb}
f^0_{D\to\pi}(\qsqmax)\underset{m_\pi^2\to 0}{\longrightarrow}\frac{f^{(D)}}{f}\qquad\textrm{and}\qquad
f^0_{B\to\pi}(\qsqmax)\underset{m_\pi^2\to 0}{\longrightarrow}\frac{f^{(B)}}{f}\,,
\end{equation}
where $f^{(D)}$ and $f^{(B)}$ are the $D$ and $B$-meson decay constants in the $\SU(2)$ chiral
limit. This relation is also valid in the static limit
($m_b\to\infty$)~\cite{Wise:1992hn,Burdman:1992gh,Wolfenstein:1992xh}
and with $1/m_b$ corrections included in $f^0_{B\to\pi}$ and
$f^{(B)}$~\cite{Burdman:1993es}. For a fixed finite value of $m_c$ or $m_b$ and
for sufficiently small $m_\pi$, the chiral corrections to the relations in (\ref{eq:callantreimanb}) are given
by eq.~(\ref{eq:f0qsqmaxchpt1}) with $f^{(K)}$ replaced by $f^{(D)}$ or $f^{(B)}$
and the low energy constants $\lambda_1$ and $\lambda_2$ also depending on
$m_c$ or $m_b$ (as well as $m_s$ through strange sea-quark effects). In the static limit on the other hand, the approach to the chiral limit becomes\,\cite{Becirevic:2002sc}
\begin{equation}\label{eq:f0qsqmaxchpt4}
f^0_{B_\textrm{static}\to\pi}(q^2_{\textrm{max}})=\frac{f^{(B_{\textrm{static}})}}{f}\left[1-(\frac{11}{4}+\frac94 g^2_{BB^\ast\pi})\,L+\frac{\lambda_{\textrm{static}}}{(4\pi f)^2}m_\pi^2+\cdots\right]\,,
\end{equation}
where $g_{BB^\ast\pi}$ is the static $BB^\ast\pi$ coupling. For fixed finite values of $m_b$,
eq.~(\ref{eq:f0qsqmaxchpt1})
may only represent the approach to the chiral limit at values of
$m_\pi$ which are smaller than those accessible in lattice simulations
and maybe even smaller than the physical value of $m_\pi$. If that is
the case then the behaviour of the form factor as a function of
$m_\pi$ will have to be studied either using the Heavy Quark Effective
Theory (or NRQCD) as was done in the static limit in
refs.~\cite{Falk:1993fr,Becirevic:2002sc} or by paying explicit
attention to the relative numerical values of $\lqcd^2/m_b$ and
$m_\pi$ in QCD calculations.

Eq.~(\ref{eq:callantreimanb}) provides an interesting check on the chiral extrapolations
of lattice calculations of leptonic decay constants and semileptonic
form factors of $B$ and $D$ mesons. As these quantities are now being
quoted with impressively small errors it is useful to have a
constraint on the extrapolations. We postpone a detailed discussion of
this issue to a future publication, but illustrate our point with an
example. In fig.~14 of ref.~\cite{Dalgic:2006dt} the authors plot the
form factors $f^+_{B\to\pi}$ and $f^0_{B\to\pi}$ as a function of
$q^2$ in the chiral limit. Superimposed on the computed points is the
fitted Ball-Zwicky parametrization~\cite{Ball:2004ye} which suggests
$f^0_{B\to\pi}(\qsqmax)\simeq 1.1$, which is very considerably below
the expected value of $f^{(B)}/f$ of greater than $1.7$ or
so\,\footnote{The points in the figure use the pre-erratum values from
  ref.~\cite{Dalgic:2006dt}. However, the changes to the results for
  $f^0_{B\to\pi}$ are small enough not to affect the statement made
  here.}. This is an interesting puzzle which remains to be resolved.
We have chosen this example because the $f^0(q^2)$ is helpfully
presented in ref.~\cite{Dalgic:2006dt} after the extrapolation to the
chiral limit has been performed and so the results were relatively
easy to interpret. A comparison of the chiral extrapolations in other
studies with eq.~(\ref{eq:callantreimanb}) remains to be undertaken.

\section{Summary and conclusions}\label{sec:concs}
In this paper we have studied the behaviour of the $K_{\ell 3}$ form factors as a function of the light quark masses ($m_u=m_d$) using $\SU(2)$ ChPT. At $q^2=0$, there is the subtlety that the final state pion is \textit{hard}, nevertheless we have shown in section \ref{sec:fqsq0} that it is possible to calculate the $\SU(2)$ chiral logarithms. The one-loop expressions are
given in eqs.~(\ref{eq:f0qsq0}) and (\ref{eq:fminusqsq0}). The coefficient of the chiral logarithm is small and we envisage that these formulae will be useful in extrapolating the results obtained for $f^0(0)$ in lattice simulations to the physical quark masses enabling a precise determination of the CKM matrix element $V_{us}$\,.

Following the procedure proposed in ref.~\cite{Becirevic:2004ya}, lattice computations of $K\to\pi$ semileptonic form factors start with a very precise determination of $f^0(\qsqmax)$. In table\,\ref{tab:prlresults} we see that the results at the values of the quark masses where the computations are performed are about 25\% below the value in the chiral limit, $f^{(K)}/f$. We investigated the chiral behaviour of $f^0(\qsqmax)$ in an attempt to understand the difference of the lattice results from the value in the $\SU(2)$ chiral limit, $f^{(K)}/f$. The coefficient of the chiral logarithms is of approximately the correct magnitude to account for this difference, but the sign is wrong; the $O(m_\pi^2\log(m_\pi^2))$ terms tend to make $f^0(\qsqmax)$ decrease as we approach the chiral limit. We estimated the coefficient of the linear and quadratic terms in $m_\pi$ using $\SU(3)$ ChPT and found large effects, which lead to an increase in $f^0(\qsqmax)$ as $m_\pi$ decreases. In this way we obtain a semi-quantitative understanding of the difference of the lattice results from $f^{(K)}/f$, but the large one-loop corrections prevent us from being able to determine the chiral behaviour with precision.

There are a number of ways in which this exploratory investigation can be improved; one natural and necessary extension would be to perform the calculations at two-loop order in ChPT. The nature of the complementary relationship between the lattice and ChPT communities is changing in a very interesting way. Until recently, lattice computations were performed with quark masses which were at best marginally in a regime where ChPT could be applied ($m_\pi\gtrsim 500\mev$) and existing ChPT calculations were used to estimate the extrapolation to physical masses. Now as lattice calculations are being performed further into the chiral region ($m_\pi\lesssim 300\mev$), it is becoming possible to use the observed dependence on the momenta and particularly on the masses to determine the LECs and to test the range of validity and precision of ChPT (see ref.~\cite{Allton:2008pn} for one such recent discussion). We stress that in order for higher-order ChPT calculations to be useful in this, the expressions should be presented in terms of mass-independent LECs and with all the mass dependence exhibited explicitly. Of course, up to now, the primary aim of ChPT calculations has been to obtain predictions for physical quantities, i.e. for quantities at physical values of the quark masses, and for this it is sufficient to express the results with the numerical values for the physical decay constants and other quantities inserted into the expressions; for such a two-loop study of $K_{\ell 3}$ decays see ref.~\cite{Bijnens:2003uy}. This prevents a determination of the full dependence on the pion masses and the calculated expressions cannot be used directly in conjunction with the lattice results.

The calculations of the $K_{\ell 3}$ form factors in ref.~\cite{Boyle:2007qe} were performed in unitary QCD, i.e. with valence and sea quark masses equal, and in this paper we studied the chiral behaviour using the standard unitary QCD. The use of partially quenched lattice simulations, in which the valence and sea quarks have different masses, is frequently a valuable tool in understanding the chiral behaviour of physical quantities. It would be a simple extension of the current work to evaluate the one-loop chiral logarithms in partially quenched QCD. Related to this is the use of partially twisted boundary conditions~\cite{Sachrajda:2004mi,Bedaque:2004ax} in order to improve the momentum resolution in the calculations of form factors in general and enabling the evaluation of $f^0(0)$ directly without an extrapolation in $q^2$~\cite{Boyle:2007wg}. The corresponding chiral and finite-volume corrections can also be evaluated.

We wish to stress the generality of the relation between the semileptonic form factor $f^0_{P\to\pi}(\qsqmax)$ and the ratio of decay constants $f^{(P)}/f$ in the $SU(2)$ chiral limit. It hold for all pseudoscalar mesons $K,\,D$ and $B$ and for sufficiently small values of $m_\pi$ the approach to the chiral limit at $\qsqmax$ is given by the simple generalization of (\ref{eq:f0qsqmaxchpt1}).
Lattice calculations of the form-factors for semileptonic $B\to\pi$ decays are performed with the pion having a small momentum, i.e. at large values of $q^2$. As usual, extrapolations in momenta and quark masses (as well as the lattice spacing) need to be performed and it is useful to use theoretical constraints to guide the extrapolations. One of these is the soft-pion relation in eq.~(\ref{eq:callantreimanb}) and its value in constraining the extrapolations remains to be investigated. Preliminary indications suggest that the conventional extrapolations lead to a value of $f^0_{B\to\pi}(\qsqmax)$ which is significantly smaller than $f^{(B)}/f$, but this will be studied systematically elsewhere.

\acknowledgments
We warmly thank Gilberto Colangelo and Juerg Gasser for helpful discussions and our colleagues from the RBC and UKQCD collaborations for many stimulating and enjoyable projects and particularly for those reported in   refs.~\cite{Boyle:2007qe} and \cite{Allton:2008pn} which led us to the study presented in this paper.

We acknowledge support from STFC Grant PP/D000211/1
and from EU contract MRTN-CT-2006-035482 (Flavianet).

\bibliographystyle{JHEP}
\bibliography{Kchpt_kl3}

\end{document}